\documentclass[aps,prl,twocolumn,showpacs,superscriptaddress,groupedaddress,amssymb,amsfonts,amsmath]{revtex4}  % for review and submission
\usepackage{graphicx}  % needed for figures

\def\jcp{J.~Comput.~Phys.}
\def\apjs{Astrophys.~J.~Supp.~Ser.}
\def\apj{Astrophys.~J.}
\def\apjl{Astrophys.~J.~Lett.}
\def\aap{Astron.~Astrophys.}
\def\mnras{Mon.~Not.~R.~Astron.~Soc.}

\def\pre{Phys.~Rev.~E}
\def\prx{Phys.~Rev.~X}
\def\prl{Phys.~Rev.~Lett.}

\def\ssr{Space Science Rev.}

\def\grl{Geophys.~Res.~Lett.}

\def\rmp{Rev.~Mod.~Phys.}

\newcommand{\pD}[2]{\frac{\partial #2}{\partial #1}}

\newcommand\bb[1]{\mbox{\boldmath{$#1$}}}
\newcommand\grad{\bb{\nabla}}
\newcommand\bcdot{\bb{\cdot}}
\newcommand\btimes{\bb{\times}}
\newcommand{\mc}[1]{\mathcal{#1}}

\newcommand{\ey}{\hat{\bb{y}}}
\newcommand{\ez}{\hat{\bb{z}}}

\begin{document}

\title{Magnetorotational Turbulence and Dynamo in a Collisionless Plasma}
\author{Matthew W.~Kunz}\email{mkunz@princeton.edu} \affiliation{Department of Astrophysical Sciences, Princeton University, 4 Ivy Lane, Princeton, NJ 08544, USA}
\affiliation{Princeton Plasma Physics Laboratory, P.O.~Box 451, Princeton, NJ 08543, USA}
\author{James M.~Stone}\affiliation{Department of Astrophysical Sciences, Princeton University, 4 Ivy Lane, Princeton, NJ 08544, USA}
\author{Eliot Quataert} \affiliation{Department of Astronomy and Theoretical Astrophysics Center, University of California, 501 Campbell Hall \#3411, Berkeley, CA 94720-3411, USA} 
\vskip 0.25cm

\date{\today}

%
% abstract
%
\begin{abstract}
We present results from the first 3D kinetic numerical simulation of magnetorotational turbulence and dynamo, using the local shearing-box model of a collisionless accretion disk. The kinetic magnetorotational instability grows from a subthermal magnetic field having zero net flux over the computational domain to generate self-sustained turbulence and outward angular-momentum transport. Significant Maxwell and Reynolds stresses are accompanied by comparable viscous stresses produced by field-aligned ion pressure anisotropy, which is regulated primarily by the mirror and ion-cyclotron instabilities through particle trapping and pitch-angle scattering. The latter endow the plasma with an effective viscosity that is biased with respect to the magnetic-field direction and spatio-temporally variable. Energy spectra suggest an Alfv\'{e}n-wave cascade at large scales and a kinetic-Alfv\'{e}n-wave cascade at small scales, with strong small-scale density fluctuations and weak non-axisymmetric density waves. Ions undergo non-thermal particle acceleration, their distribution accurately described by a kappa distribution. These results have implications for the properties of low-collisionality accretion flows, such as that near the black hole at the Galactic center.
\end{abstract}

\maketitle

%
% introduction
%
{\it Introduction.}---The theory of black-hole accretion is central to many areas of theoretical, computational, and observational astronomy. Not only does accretion power some of the phenomenologically richest electromagnetic sources in the Universe, but also black-hole accretion flows serve as excellent laboratories for the study of basic plasma physics and strong-field general relativity (GR). 

Recently, much attention has been paid to the latter \cite{dhk03,gmt03,hirose04,devilliers05,khh05,hk06,fragile07,nkh09,penna10,shiokawa12,narayan12,mtb12,sadowski13}, with myriad computational efforts seeking to connect the properties of simulated black-hole accretion flows in curved spacetime with the observed mm/sub-mm emission \cite{moscibrodzka09,spm12,dexter12,drappeau13,moscibrodzka14,chan15}. While fruitful, these calculations suffer from {\it ad hoc} assumptions about the nature of the accreting plasma, which is often so hot and diffuse that the collisional mean free path is comparable to (or even larger than) the system size and many orders of magnitude larger than the particles' Larmor radii. This hierarchy of scales precludes a straightforward application of the oft-employed magnetohydrodynamic (MHD) equations, and instead warrants a kinetic approach.

As a compliment to these studies, we forego any treatment of GR and instead focus on the complex interplay between micro-scale plasma processes and macro-scale dynamics. Our starting point is the magnetorotational instability (MRI; \cite{bh98}), which two decades worth of MHD simulations have shown enables mass accretion by efficiently transporting angular momentum outwards in the disk. In a weakly collisional plasma, conservation of particles' adiabatic invariants during magnetic-field amplification by the MRI and/or the Keplerian shear renders the gas pressure anisotropic with respect to the magnetic field \cite{qdh02}. On large scales, this ``pressure anisotropy'' impacts viscous heating and dynamo behavior, and can even transport as much angular momentum as the Reynolds and Maxwell stresses \cite{sharma06}. On small scales, this anisotropy drives high-frequency waves and kinetic microinstabilities (e.g., firehose, mirror), which provide an enhanced rate of particle scattering and affect the topology of the magnetic field \cite{kss14,riquelme15}. The magnetic Prandtl number ${\rm Pm}$, known to be important for the saturation of the MRI \cite{ll07,fromang07,sh09}, thus becomes a dynamical quantity set by wave-particle interactions.

To elucidate the impact of these processes on collisionless accretion, we present results from the first 3D kinetic simulation of magnetorotational turbulence and dynamo. This follows several recent papers on the linear stability of collisionless accretion disks \cite{qdh02,hq14,quataert15} and the nonlinear evolution of 2D kinetic magnetorotational turbulence \cite{riquelme12,hoshino13,ksb14}, as well as one paper on the 3D nonlinear evolution of a kinetic-MRI ``channel'' mode in a pair plasma \cite{hoshino15}. Our work also provides an {\em ab initio} kinetic foundation for recent efforts to include kinetic effects into the equations of GRMHD for studies of black-hole accretion \cite{chandra15,foucart16}, as well as for the pioneering simulations of magnetorotational turbulence in a collisionless plasma by Sharma {\it et al.}~\cite{sharma06}, who used kinetic-MHD equations with a Landau-fluid closure and pressure-anisotropy limiters.

%
% Equations
%
{\it Hybrid-kinetic equations in the shearing box.}---We consider a differentially rotating (Keplerian) disk of non-relativistic, quasi-neutral, collisionless, and initially homogeneous plasma of electrons (mass $m_e$, charge $-e$) and ions (mass $m_i$, charge $e$) threaded by a magnetic field. In a local Cartesian $(x,y,z)$ frame comoving with the disk and centered at a fiducial radial location $r_0$---the ``shearing box'' \cite{gl65,hgb95}---the equations governing the evolution of the ion distribution function $f_i(t, \bb{r}, \bb{v})$ and the magnetic field $\bb{B}(t,\bb{r})$ are, respectively, the Vlasov equation
\begin{align}\label{eqn:vlasov}
\biggl( \pD{t}{} &- \frac{3}{2}\Omega_{\rm rot} x \pD{y}{} \biggr) f_i + \bb{v} \bcdot \grad f_i + \frac{3}{2} \Omega_{\rm rot} v_x \pD{v_y}{f_i} \nonumber\\*
\mbox{} &+ \biggl[ \frac{e}{m_i} \Bigl( \bb{E}' + \frac{\bb{v}}{c} \btimes \bb{B} \Bigr) - 2\Omega_{\rm rot} \ez\btimes\bb{v} \biggr] \bcdot \pD{\bb{v}}{f_i}  = 0
\end{align}
and Faraday's law
\begin{equation}\label{eqn:induction}
\biggl( \pD{t}{} - \frac{3}{2}\Omega_{\rm rot} x \pD{y}{} \biggr) \bb{B} = - c \grad \btimes \bb{E}' - \frac{3}{2}\Omega_{\rm rot} B_x \ey ,
\end{equation}
where $\bb{\Omega}_{\rm rot} = \Omega_{\rm rot}\ez$ is the angular velocity at $r_0$. The $x$ and $y$ dimensions coincide locally with the radial and azimuthal dimensions in the disk. The electric field in the comoving frame
\begin{equation}\label{eqn:efield}
\bb{E}' = - \frac{\bb{u}_i \btimes \bb{B}}{c} + \frac{\bb{j} \btimes \bb{B}}{ce n_i} - \frac{T_e \grad n_i}{e n_i} + \frac{4\pi\eta}{c^2} \bb{j},
\end{equation}
is obtained by expanding the electron momentum equation in $( m_e / m_i )^{1/2}$, enforcing quasi-neutrality
\begin{equation}\label{eqn:quasineutrality}
n_e = n_i \equiv \int {\rm d}^3 \bb{v} \, f_i ,
\end{equation}
assuming isothermal electrons ($T_e = {\rm const.}$), and using Amp\'{e}re's law to solve for the mean electron velocity
\begin{equation}\label{eqn:ampere}
\bb{u}_e = \bb{u}_i - \frac{\bb{j}}{e n_i} \equiv  \frac{1}{n_i} \int {\rm d}^3 \bb{v} \, \bb{v} f_i - \frac{ c \grad \btimes \bb{B} }{4 \pi e n_i} 
\end{equation}
in terms of the mean ion velocity $\bb{u}_i$ and the current density $\bb{j}$ \cite{bcch78,hn78}. A resistivity $\eta$ is included in (\ref{eqn:efield}) to remove small-scale magnetic energy. Eqs.~(\ref{eqn:vlasov})--(\ref{eqn:ampere}) constitute the ``hybrid'' description of kinetic ions and fluid electrons \cite{hn78,winske85,lipatov02,wyokq03}, tailored for the unstratified shearing box \cite{ksb14,hq14}.

%
% method of solution
%
{\it Method of solution.}---We solve Eqs.~(\ref{eqn:vlasov})--(\ref{eqn:ampere}) using the second-order--accurate particle-in-cell code {\sc Pegasus} \cite{ksb14}. $N_p = 64 N_x N_y N_z$ ion particles are drawn from a Maxwell distribution with $\beta_{i0} \equiv v^2_{thi0}/v^2_{A0} = 200$ and placed on a 3D shearing-periodic grid with $N_x \times N_y \times N_z = 384 \times 1536 \times 384$ cells spanning $L_x \times L_y \times L_z = H \times 4H \times H$, where $H \equiv v_{thi0} / \Omega_{\rm rot}$ is the disk scale height, $v_{thi0} \equiv (2T_{0i}/m_i)^{1/2}$ is the ion thermal speed, and $v_{A0} \equiv B_0/(4\pi m_i n_{0i})^{1/2}$ is the Alfv\'{e}n speed; the subscript ``0'' denotes an initial value. We assume zero mean magnetic flux: initially, $\bb{B}_0 = B_0 \sin(2\pi x/H) \ez$. If amplified and sustained by the MRI, this field configuration would constitute a ``magnetorotational dynamo'' \cite{hgb96}. The initial ion gyrofrequency $\Omega_{i0} \equiv eB_0/m_ic = 50\Omega_{\rm rot}$; the initial ion Larmor radius $\rho_{i0} \equiv v_{thi0} / \Omega_{i0} = 0.02H$. The electrons are Maxwellian and isothermal with $T_e = T_{i0}$, so that the total initial plasma $\beta_0 = \beta_{i0} + \beta_{e0} = 400$. The magnetic Reynolds number ${\rm Rm} \equiv \Omega_{\rm rot} H^2 / \eta = 37,500$. These parameters provide reasonable scale separation between the grid scale, the Larmor scale, and the box size, one which improves as the MRI grows and the plasma becomes more magnetized. The moments $n_i$ and $n_i\bb{u}_i$ are low-pass filtered once per timestep to mitigate feedback from finite-particle-number noise. A fourth-order hyper-resistivity is used to damp dispersive fluctuations at the grid. In what follows, $\langle \cdot \rangle$ denotes a spatial average; $\langle\!\langle\cdot\rangle\!\rangle$ denotes a spatio-temporal average.

%
% figure 1
%
\begin{figure}
\centering
\includegraphics[width=3.375in,clip]{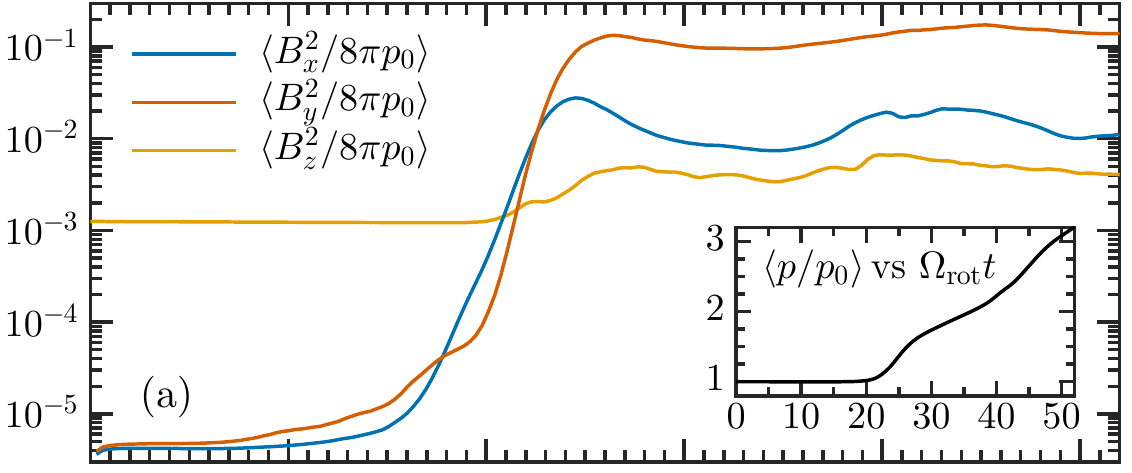}
\newline
\includegraphics[width=3.375in,clip]{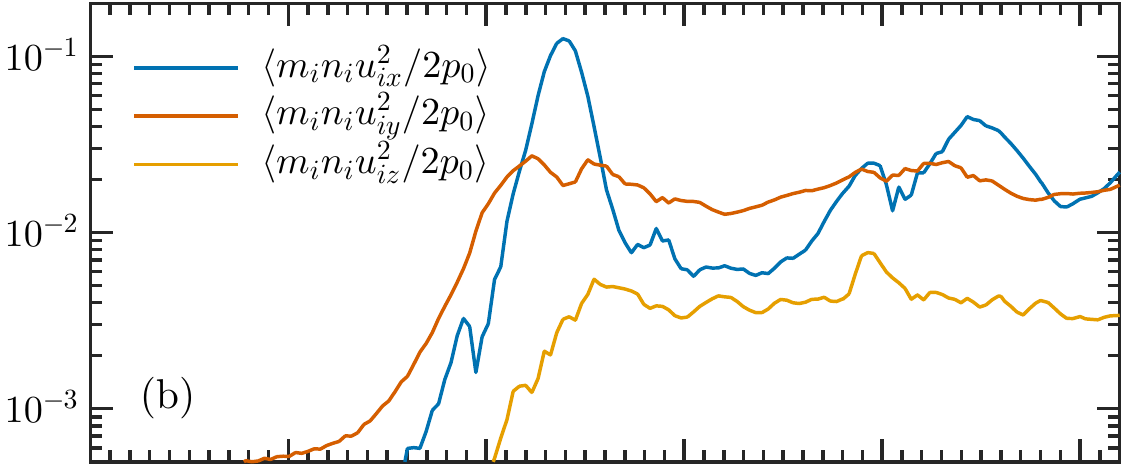}
\newline
\includegraphics[width=3.375in,clip]{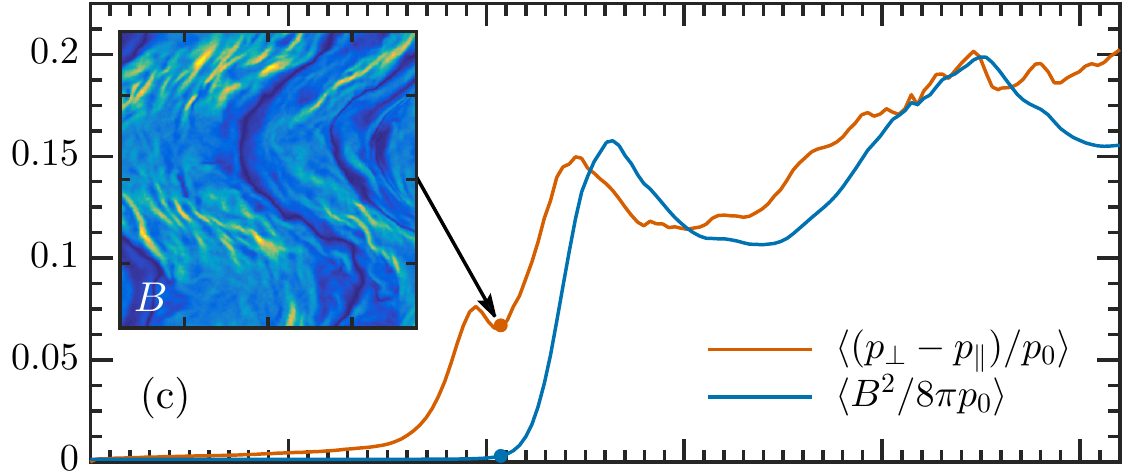}
\newline
\includegraphics[width=3.375in,clip]{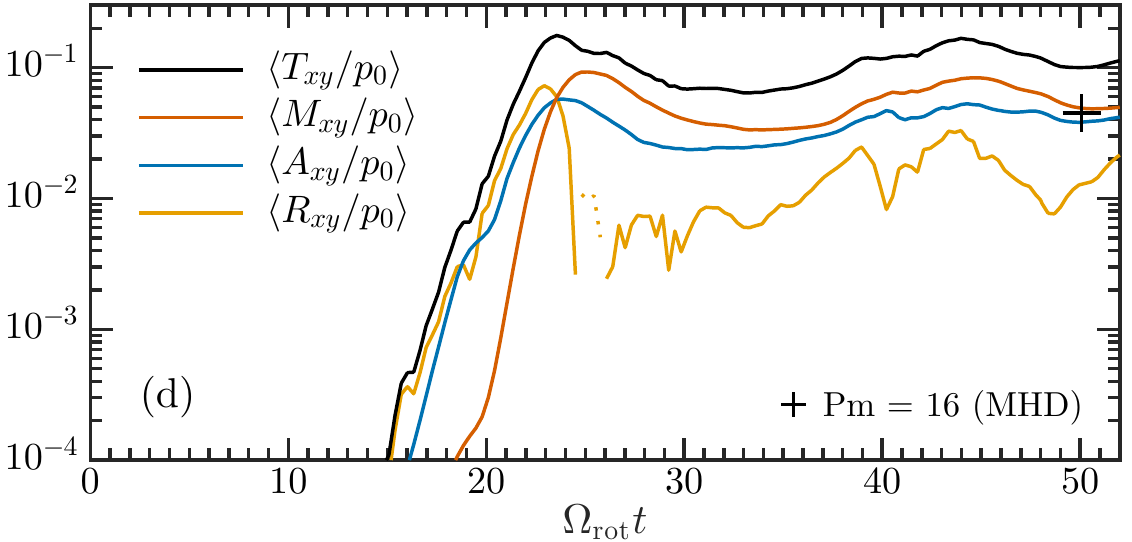}
\newline
\vspace{-0.6cm}
\caption{Evolution of box-averaged (a) magnetic energy and thermal pressure, (b) kinetic energy, (c) pressure anisotropy (compared to magnetic energy), and (d) $xy$ components of the total, Maxwell, viscous, and Reynolds stresses, all normalized to initial thermal pressure $p_0$. The inset in (c) shows a slice of the magnetic-field strength in the $x$-$z$ plane at the time marked by the dot; mirror-mode parasites, which feed off the pressure anisotropy generated by the MRI, are evident. The plus sign in (d) denotes the value of $\langle\!\langle T_{xy}/p_0\rangle\!\rangle$ obtained in an MHD simulation of the zero-net-flux MRI with ${\rm Pm} = 16$ \cite{fromang07}.}
\label{fig:mri-energy}
\end{figure}

\begin{figure}
\centering
\includegraphics[width=3.375in,clip]{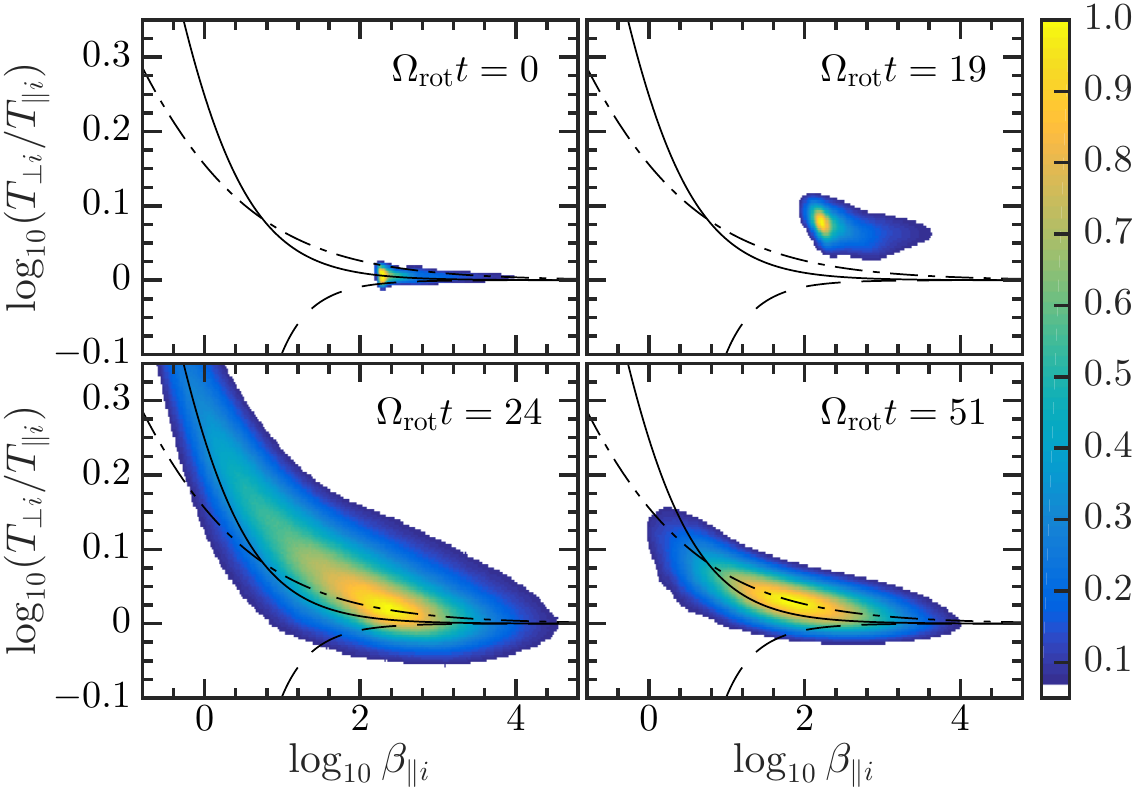}
\newline
\vspace{-0.6cm}
\caption{Distribution of ion temperature anisotropy $T_{\perp i}/T_{\parallel i}$ versus parallel ion beta $\beta_{\parallel i}$ (top left) initially, (top right) during the channel phase, (bottom left) at peak channel amplitude, and (bottom right) in the saturated state. The solid, dot-dashed, and dashed lines denote approximate mirror, ion-cyclotron, and firehose instability thresholds, respectively.}
\label{fig:mri-bale}
\end{figure}

%
% results
%
{\it Results.}---Figure \ref{fig:mri-energy}(a) presents the evolution of the box-averaged magnetic and thermal pressures. In the early, linear (``channel'') phase, the MRI grows the horizontal components of the magnetic field exponentially. By adiabatic invariance, this produces pressure anisotropy [Fig.~\ref{fig:mri-energy}(c)], with $\langle p_\perp \rangle > \langle p_\parallel \rangle$. This anisotropy affects the evolution of the MRI in three ways. First, it pushes the instability to longer wavelengths by supplementing the magnetic tension. Secondly, it provides a free-energy source for ion-Larmor-scale mirror-mode parasites, some of which can be seen in Fig.~\ref{fig:mri-energy}(c)-inset. These modes reduce the pressure anisotropy, ultimately limiting it to be comparable to the box-averaged magnetic pressure \footnote{In a more realistic model with larger $\Omega_i/\Omega_{\rm rot}$, the mirror instability would grow rapidly enough to efficiently regulate the pressure anisotropy. Instead, with $\Omega_i / \Omega_{\rm rot} \simeq 50$ in the channel phase, the pressure anisotropy significantly overshoots the mirror threshold before being regulated. Even in the saturated state, when $\Omega_i/\Omega_{\rm rot} \simeq 500$, regulation is not perfect (see Fig.~\ref{fig:mri-bale}). Dedicated studies of the mirror instability \cite{kss14} suggest $\Omega_i/\Omega_{\rm rot} \gtrsim$$10^3$ is needed to achieve asymptotic behavior. Since the computational cost $\propto$$(\Omega_{i0}/\Omega_{\rm rot})^4$ at fixed Larmor-scale resolution, doing substantially better is not currently feasible.}. Finally, pressure anisotropy generates a ``viscous'' stress ($A_{xy}$), which supplements the angular-momentum transport customarily afforded by the Reynolds ($R_{xy}$) and Maxwell ($M_{xy}$) stresses:
\begin{align}
T_{xy} &= R_{xy} + M_{xy} + A_{xy} \nonumber\\*
\mbox{} &\equiv m_i n_i u_x u_y - \frac{B_x B_y}{4\pi} - ( p_\perp - p_\parallel ) \frac{B_x B_y}{B^2} .
\end{align}
These stresses are shown, box averaged, in Fig.~\ref{fig:mri-energy}(d). At $\Omega_{\rm rot} t \approx 25$, the channel breaks down into magnetorotational turbulence, with the magnetic energy dominated by its azimuthal component [Fig.~\ref{fig:mri-energy}(a)], the kinetic energy being comparable to the magnetic energy [Fig.~\ref{fig:mri-energy}(b)], the pressure anisotropy regulated by the mirror instability to be comparable to the magnetic pressure [Fig.~\ref{fig:mri-energy}(c)], and the viscous and Maxwell stresses supplying most of the angular-momentum transport [Fig.~\ref{fig:mri-energy}(d)], with $\alpha \doteq \langle\!\langle T_{xy}/p_0\rangle\!\rangle \sim 0.1$. With Keplerian rotation enforced by the shearing boundaries, this stress does work on the plasma and heats it continuously [Fig.~\ref{fig:mri-energy}(c)-inset].

%
% figure 2
%

Fig.~\ref{fig:mri-bale} shows the data distribution in the $(T_{\perp i}/T_{\parallel i})$-$\beta_{\parallel i}$ plane at four times. Approximate thresholds for mirror, ion-cyclotron, and firehose instabilities are from Ref.~\cite{hellinger06} (assuming bi-Maxwellian ions and Maxwellian $\beta_e = 1$ electrons). Initially (top left), the ion distribution is isotropic, with $\beta_{\parallel i} \approx 200$ (the tail extending to higher $\beta_{\parallel i}$ is due to the zero-net-flux configuration). As the MRI exponentially amplifies the magnetic-field strength, adiabatic invariance drives $T_{\perp i} > T_{\parallel i}$ (top right), lifting the distribution upwards beyond the mirror and ion-cyclotron thresholds. Thereafter, mirror-mode parasites isotropize the distribution to lie close to the mirror threshold, along which it runs to smaller (larger) $\beta_{\parallel i}$ ($T_{\perp i}/T_{\parallel i}$) (bottom left, at peak channel amplitude). As the channel breaks down into turbulence, the distribution settles into a configuration with minimum $\beta_{\parallel i} \sim 1$, constrained near the mirror threshold at high $\beta_{\parallel i}$ and the ion-cyclotron threshold at low $\beta_{\parallel i}$ (bottom right). The propensity for the MRI to amplify the magnetic field and thus drive $T_{\perp i} > T_{\parallel i}$ means that very little of the plasma lies at the firehose threshold.

\begin{figure}
\centering
%fld output #151; mom output #112
\includegraphics[width=3.375in,clip]{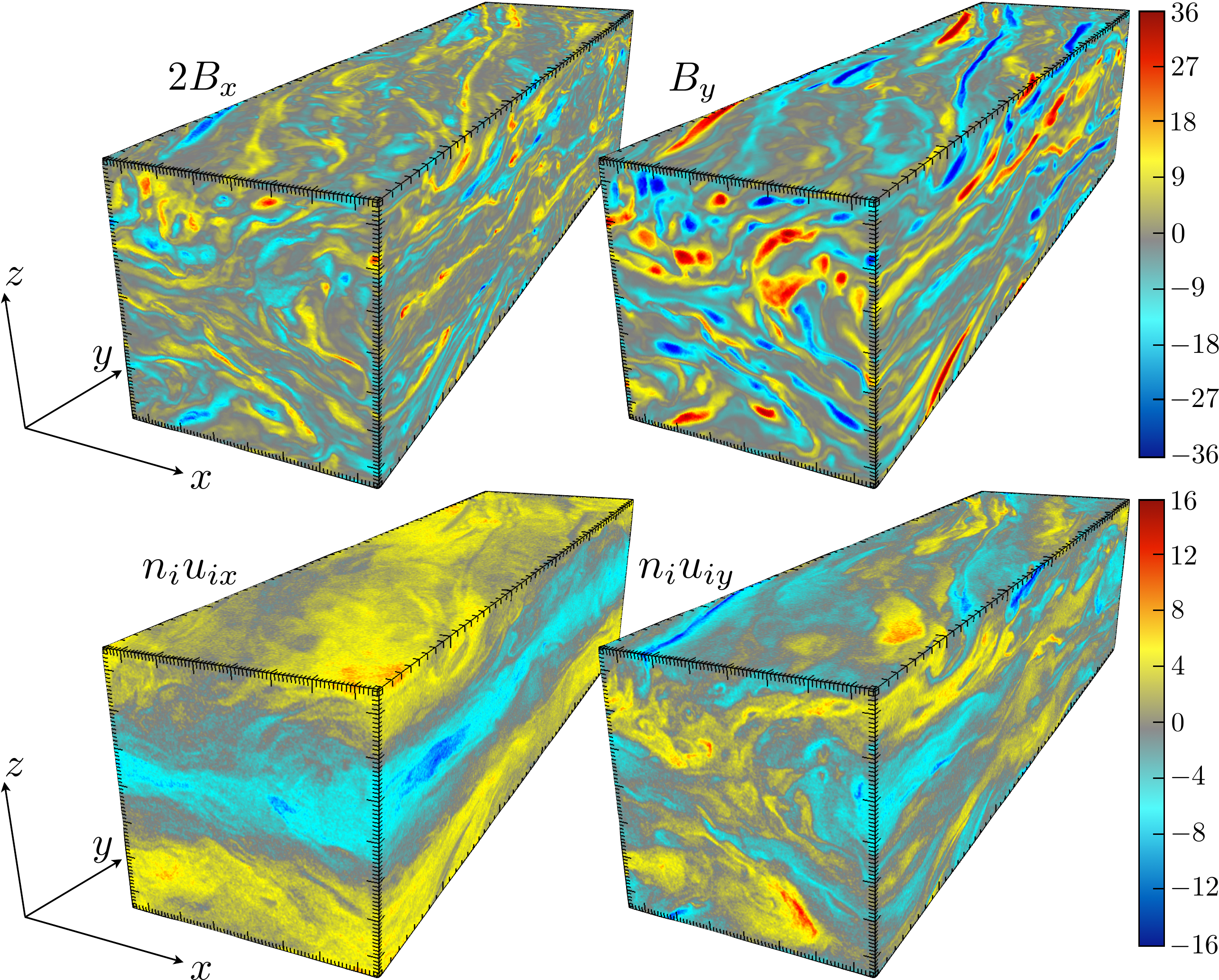}
\newline
\vspace{-0.5cm}
\caption{$x$ and $y$ components of the magnetic field (normalized to $B_0$) and momentum (normalized to $n_{i0} v_{A0}$) at $\Omega_{\rm rot} t = 47.4$.}
\label{fig:mri-fields}
\end{figure}

Figure \ref{fig:mri-fields} displays pseudo-color images of the magnetic-field and momentum fluctuations at $\Omega_{\rm rot} t = 47.4$. The magnetic flux is arranged into thin, azimuthally extended bundles with short perpendicular scales, separated by patches of small-scale turbulence, all with $B_x$ and $B_y$ anti-correlated. The momentum appears larger in scale, with large swathes being comparatively laminar (especially in the $x$ component). This is a clear example of a collisionless, magnetized, high-$\beta$ plasma behaving as though it were a large-Pm fluid (albeit with stifled cross-field viscosity due to the small Larmor radii). 

\begin{figure}
\centering
%fld output #151, mom output #112
\includegraphics[width=3.375in,clip]{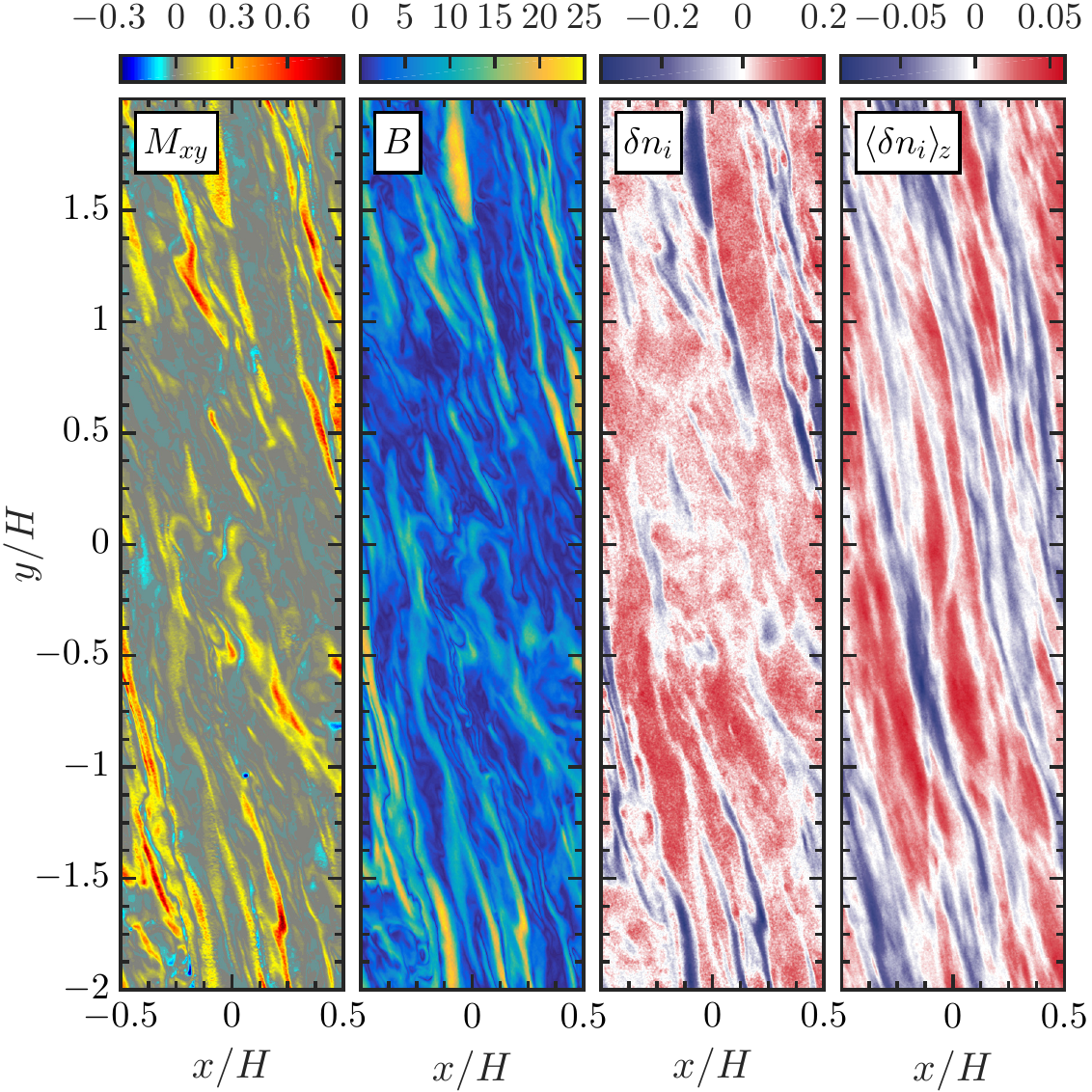}
\vspace{-0.5cm}
\caption{Slices of (left) Maxwell stress (normalized to $p_0$), (left-center) magnetic-field strength (normalized to $B_0$), and (right-center) density fluctuation (normalized to $n_{0i}$) at $z=0$. (right) Vertically averaged density fluctuation (normalized to $n_{0i}$). All frames taken at $\Omega_{\rm rot} t = 47.4$.}
\label{fig:mri-slices}
\end{figure}

Slices of the computational domain at $z=0$ showing the Maxwell stress $M_{xy}$, the magnetic-field strength $B$, and the perturbed density $\delta n_i \equiv n_i - \langle n_i \rangle$ are given in the leftmost two panels of Fig.~\ref{fig:mri-slices}. The Maxwell stress is largest in thin, azimuthally extended filaments, separated by wide regions of almost zero stress (cf.~fig.~4 of \cite{bodo11}). The field strength is largely anti-correlated with the density fluctuations, particularly in small-scale mirrors (where particles congregate in regions of weak field) and in large-scale bundles of compressed magnetic field (from which particles have been largely evacuated). Such large density fluctuations ($\gtrsim$$10\%$) on these scales are not seen in comparable MHD runs. The prominent $k_z=0$ non-axisymmetric density waves seen in compressible MHD simulations of magnetorotational turbulence are absent here. Only after integrating over height (rightmost panel) do non-axisymmetric density waves appear, and then only at relatively small amplitudes (compare to figs 2 and 3 of Ref.~\cite{hp09b}). This may be due to strong Landau damping of sound waves, a feature absent in MHD.

%
% figure 4
%

%
% figure 5
%
\begin{figure}
\centering
\includegraphics[width=3.375in,clip]{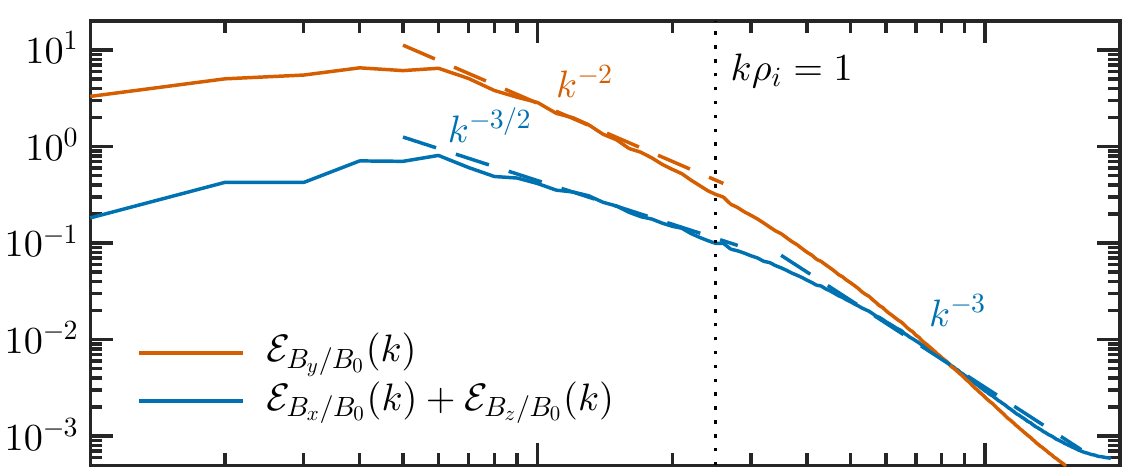}
\newline
\includegraphics[width=3.375in,clip]{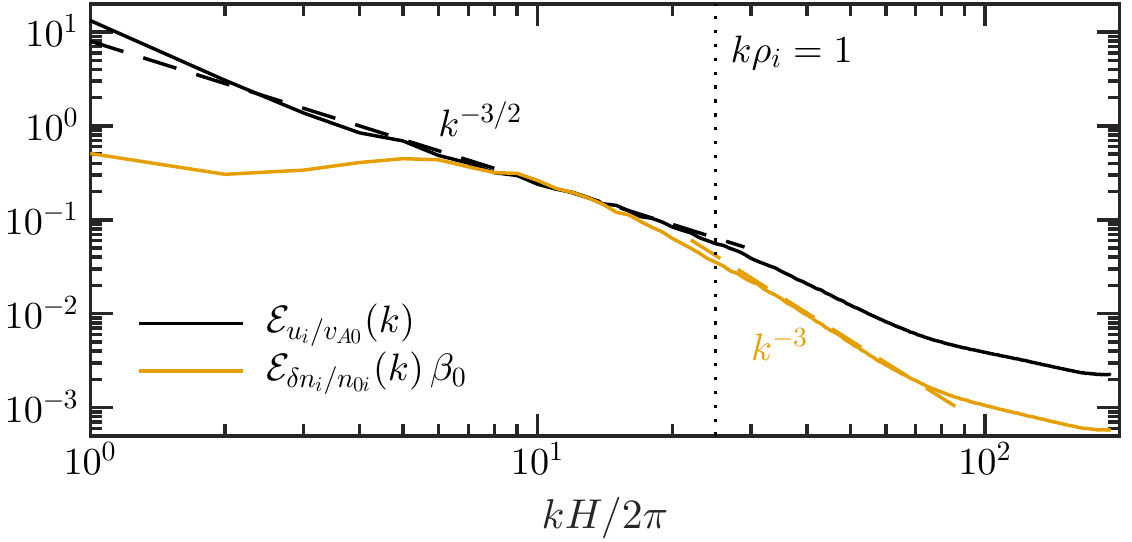}
\newline
\vspace{-0.5cm}
\caption{Energy spectra of (top) magnetic fluctuations and (bottom) velocity and density fluctuations in the saturated state, defined by $\mc{E}_A(k) \equiv \int{\rm d}\Omega_k \,(kH/2\pi)^2 |A_k|^2$ with $\int{\rm d}(kH/2\pi) \, \mc{E}_A(k) = \langle A^2\rangle$. Characteristic slopes are shown as labelled dashed lines; the vertical dotted line marks $k\rho_i = 1$.}
\label{fig:mri-spectra}
\end{figure}

Energy spectra of the magnetic-field, ion-velocity, and density fluctuations in the turbulent saturated state are given in Fig.~\ref{fig:mri-spectra}. Above ion-Larmor scales ($k \rho_i < 1$), the kinetic and poloidal magnetic spectrum vary as $k^{-3/2}$, while the azimuthal magnetic energy $\propto$$k^{-2}$. These spectra resemble those obtained in recent high-resolution incompressible MHD simulations of the MRI \cite{wlb16}. By analogy with the $k^{-3/2}$ spectrum that is almost universally obtained within the inertial range of driven, strong MHD turbulence with a guide field \cite{mg01,mcb06,mcb08,chen11,perez12,csm15}, the spectra in Fig.~\ref{fig:mri-spectra} can be viewed as describing small-scale Alfv\'{e}nic turbulence guided locally by a large-scale, predominantly azimuthal field, whose $k^{-2}$ spectrum is likely due to sharp field-direction reversals at the boundaries of otherwise coherent magnetic domains \cite{wlb16}. (Mirror instability is predicted to produce a power-law spectrum $\propto$$k^{-5/3}$ at $k \rho_i \lesssim 1$ \cite{kss14}, but with amplitudes too small to easily distinguish in the spectrum.) Note the deficit of density fluctuations at long wavelengths. At sub-ion-Larmor scales ($k \rho_i > 1$), the density and magnetic spectra steepen to take on a slope ($k^{-3}$) and polarization ($\delta n_i \sim \beta^{-1}_i \delta B$) characteristic of kinetic-Alfv\'{e}n-wave turbulence \cite{schekochihin09,bp12,ps15}. This marks the first time that such a cascade has been observed in magnetorotational turbulence, and suggests that certain aspects of gyrokinetic \cite{schekochihin09,howes11} and solar-wind turbulence \cite{alexandrova09,sahraoui10,salem12,chen13} may be useful for understanding dissipation in collisionless accretion disks (e.g., \cite{howes10}).

\begin{figure}
\centering
%from rst.0032
\includegraphics[width=3.375in,clip]{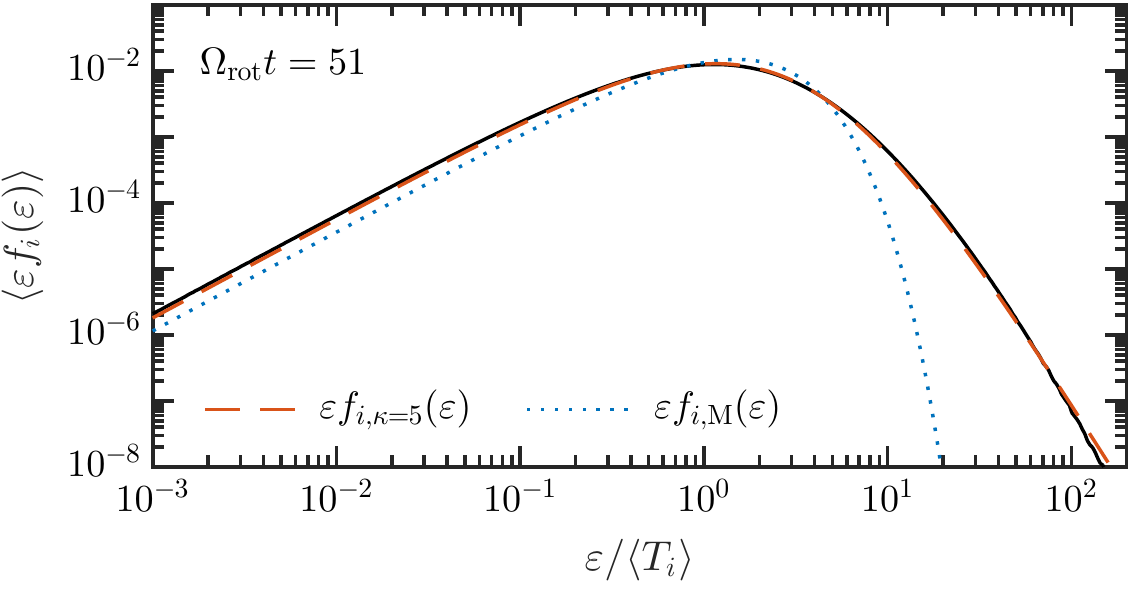}
\vspace{-0.6cm}
\caption{Box-averaged ion distribution function at $\Omega_{\rm rot} t = 51$ (solid line), binned logarithmically in $\varepsilon \equiv (m_i/2) | \bb{v} - \bb{u}_i(\bb{r}) |^2$. A $\kappa = 5$ distribution and a Maxwell distribution, both with temperature $\langle T_i\rangle \simeq 5.4T_{0i}$, are overlaid; the former is a good fit, indicating non-thermal particle acceleration.}
\label{fig:mri-distfunc}
\end{figure}

Finally, Fig.~\ref{fig:mri-distfunc} presents the ion distribution function at the end of the run versus $\varepsilon \equiv (m_i/2)|\bb{v}-\bb{u}_i(\bb{r})|^2$, the particle energy measured in the frame of the local mean ion velocity. A Maxwell distribution $f_{{\rm M}}(\varepsilon) \propto \sqrt{\varepsilon} \exp(-\varepsilon/ T)$ and a kappa distribution $f_{\kappa}(\varepsilon) \propto \sqrt{\varepsilon} \,[ 1 + (\varepsilon / T)/ (\kappa-3/2) ]^{-(\kappa+1)}$ with $\kappa = 5$ are provided for reference, with $T = \langle T_i \rangle \equiv (2/3) \langle ( \int{\rm d}\varepsilon\, \varepsilon f_i ) / ( \int{\rm d}\varepsilon\, f_i ) \rangle \simeq 5.4T_{0i}$. The distribution function is clearly non-thermal, with $f_{\kappa = 5}$ being a good fit (although $\kappa$ is likely still decreasing).

%
% figure 6
%

%
% summary
%
{\it Summary.}---Many of the gross qualitative features of the turbulence found here are reminiscent of those obtained in MHD simulations. These include correlated fluctuations leading to efficient outward angular-momentum transport, amplification and sustenance of a subthermal magnetic field, azimuthally biased magnetic-field direction, and some aspects of the energy spectra. Given that strong particle-particle collisions have been replaced here by wave-particle interactions, this resemblance is notable, and lends hope to the idea that fluid models of collisionless, magnetized plasmas might suffice in describing much of the macroscale evolution. 

But there are important differences, mostly due to the allowed departures of the ion distribution function from an isotropic Maxwellian. These departures, driven by adiabatic invariance and shaped by the local magnetic-field direction, produce additional angular-momentum transport and generate ion-Larmor-scale fluctuations that trap and pitch-angle scatter particles. The latter endow the plasma with a large (but highly anisotropic and spatially variable) magnetic Prandtl number. As a result, the magnetic-field geometry is dominated by thin, azimuthally elongated flux tubes with short perpendicular dimension. The velocity is relatively laminar, with coherent large-scale features that persist over several orbits. Other notable features include the relatively weak excitation of non-axisymmetric density waves (as compared to MHD), the strong density inhomogeneities on small scales, the development of a sub-ion-Larmor kinetic-Alfv\'{e}n-wave cascade, and a broad ion distribution function indicative of non-thermal particle acceleration. 

Our assumption of isothermal, Maxwellian electrons makes comparison with observations difficult since electrons dominate the emission. Electrons and ions are expected to be heated differently depending upon local plasma conditions \cite{qg99,sharma07,cranmer09,howes10,sn15,sironi15}, a feature that plays a defining role in several theories of black-hole accretion \cite{rees82,ny95,nmq98}. Studying this requires a more sophisticated treatment of electron thermodynamics than in our hybrid model. In the meantime, our results provide {\it ab initio} evidence that enhanced angular-momentum transport and non-thermal particle acceleration in collisionless accretion disks is facilitated by the kinetic MRI.

%
% acknowledgments
%
\vspace{0.1in}
\begin{acknowledgments}
Support for M.~W.~K.~during the early stages of this project was provided by a Lyman Spitzer, Jr.~Fellowship. J.~M.~S.~was supported in part by NSF grant AST-1333091. E.~Q.~was supported in part by NSF grant AST 13-33612, a Simons Investigator Award from the Simons Foundation, and the David and Lucile Packard Foundation. The results of this research have been achieved using the PRACE Research Infrastructure resource Curie based in France at CEA (TGCC). This work benefitted from useful conversations with Sebastien Fromang, Greg Hammett, Tobias Heinemann, Geoffroy Lesur, Alexander Schekochihin, and Jonathan Squire. Aspects of this work were facilitated by the Max-Planck/Princeton Center for Plasma Physics (NSF grant PHY-1144374), the NSF Theoretical and Computational Astrophysics Network on Black Hole Accretion, and the Wolfgang Pauli Institute Vienna.
\end{acknowledgments}

\end{document}